\documentclass[conference]{IEEEtran}
\IEEEoverridecommandlockouts

\usepackage{cite}
\usepackage{amsmath,amssymb,amsfonts}
\usepackage{algorithmic}
\usepackage{graphicx}
\usepackage{textcomp}
\usepackage{xcolor}
\usepackage{float}
\usepackage{subfig} 
\usepackage{caption}
\usepackage{tabularx}
\usepackage{makecell}
\usepackage{multirow} 
\usepackage{tikz}
\usepackage{booktabs} 

\def\BibTeX{{\rm B\kern-.05em{\sc i\kern-.025em b}\kern-.08em
    T\kern-.1667em\lower.7ex\hbox{E}\kern-.125emX}}
\begin{document}

\title{A Hybrid BLE--Wi-Fi Communication Architecture for Adaptive Imaging in Wireless Capsule Endoscopy
\thanks{This work was supported by the Nanyang Professorship, the MOE Tier 1 grant RG71/24, and the MTC MedTech Programmatic Fund M24N9b0125. \textit{(Ziyao Zhou and Zhuoran Sun contributed equally to this work.)}}}

\author{
Ziyao Zhou$^{1}$, 
Zhuoran Sun$^{1}$, 
Chen Shen$^{1}$, 
Xinyi Shen$^{1}$, 
Zhehao Lu$^{1}$, 
Sikkandar$^{1}$, 
Hen-Wei Huang$^{1,2}$\\
\IEEEauthorblockA{$^{1}$School of Electrical and Electronic Engineering, Nanyang Technological University, Singapore}
\IEEEauthorblockA{$^{2}$LKC School of Medicine, Nanyang Technological University, Singapore}
}



\maketitle

\begin{abstract}
Wireless capsule endoscopy (WCE) is fundamentally constrained by limited wireless bandwidth, resulting in low imaging resolution and frame rate that can lead to motion blur and missed lesions. Although adaptive frame-rate schemes have been explored to accommodate transient gastrointestinal (GI) motility, these approaches typically require sacrificing image resolution. The use of higher-frequency communication bands has also been limited by increased tissue attenuation.
To address these challenges, we propose a hybrid Bluetooth Low Energy (BLE)–Wi-Fi communication architecture that combines the low-power operation of BLE with the high data throughput of Wi-Fi. We systematically evaluate the performance of BLE and Wi-Fi under tissue-mimicking conditions by measuring throughput, received signal strength indicator (RSSI), and power consumption. Our results demonstrate that amplified BLE employing an adaptive transmission power (TXP) control strategy can provide a stable frame rate with low power consumption, while 2.4-GHz Wi-Fi operating in station mode emerges as the most suitable high-throughput communication configuration for WCE. Compared with Wi-Fi, BLE reduces power consumption by approximately 10×, whereas Wi-Fi achieves up to 10× higher throughput. To reconcile these complementary trade-offs, we further introduce a hybrid system using frame-boundary–synchronized switching mechanism to ensure lossless data transmission during BLE–Wi-Fi transitions. Finally, we observe that the image payload size significantly affects switching efficiency between communication modes. Experimental results show that the BLE$\rightarrow$Wi-Fi switching latency is approximately 92.66 ms, which is longer than the Wi-Fi$\rightarrow$BLE switching latency of 15.49 ms when transmitting 10 KB image payloads. Overall, the proposed hybrid BLE–Wi-Fi system enables robust, lossless, and energy-efficient mode switching, supports adaptive imaging, and advances the development of next-generation autonomous WCE platforms.

\end{abstract}


\section{Introduction}
\begin{figure}[t]
 \centering
\includegraphics[width=1\linewidth]{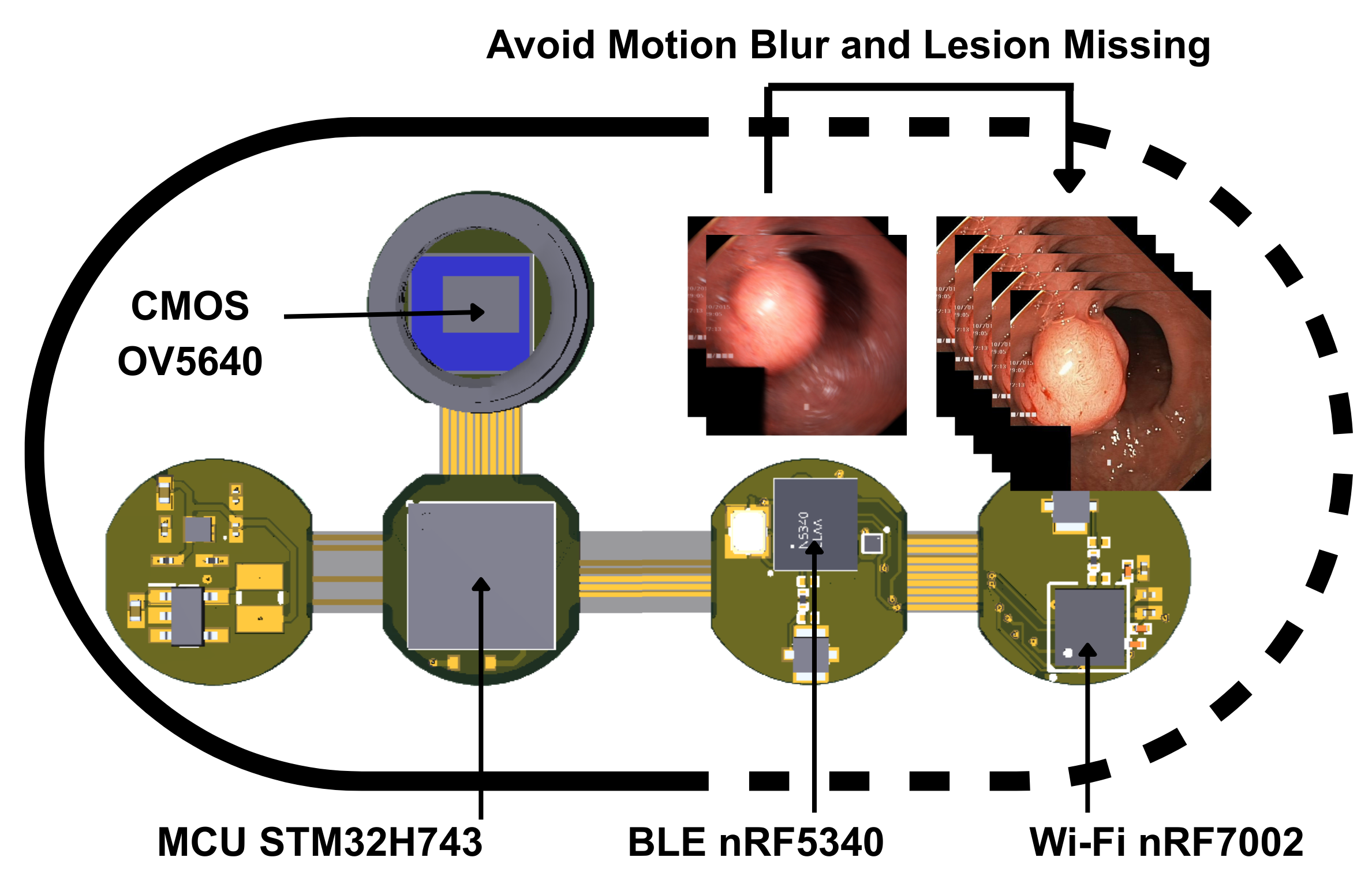}
\caption{Hybrid BLE--Wi-Fi communication architecture for wireless capsule endoscopy, enabling energy-efficient operation with lossless switching between low-frame-rate BLE transmission and high-frame-rate Wi-Fi image streaming.
}
\label{fig_SystemPCB}
\end{figure}

Wireless capsule endoscopy (WCE) remains the only truly non-invasive modality for direct visualization of the gastrointestinal (GI) tract \cite{Cao2024RoboticWCE}. Since its clinical introduction in 2001, however, improvements in imaging resolution and frame rate have been limited \cite{Cao2024Robotic}. Current commercial systems typically provide spatial resolutions below 500 × 500 pixels and frame rates of less than 5 frames per second (FPS), which can result in motion blur and an increased risk of missed lesions. Although adaptive frame-rate strategies have been introduced to temporarily increase FPS under rapid capsule motion, such approaches generally require sacrificing image quality \cite{Monteiro2016PillCam}. These limitations are primarily imposed by the restricted wireless communication bandwidth available to ingestible devices.

Most existing WCE systems operate in sub-GHz frequency bands to mitigate electromagnetic attenuation in biological tissues \cite{Zhou2025InVivo}. While this choice improves signal penetration, it inherently limits achievable data throughput. For example, the SI4455 transceiver supports data rates up to 500 kbps \cite{SiliconLabs_Si4455_2025}, and the ZL70103 operating at 402 MHz achieves a maximum of 600 kbps \cite{8743718}. One exception is the CC1310, which nominally advertises support for a 4M PHY \cite{TI_CC1310_Datasheet}. Under throughput constraints, practical WCE systems can not reliably capture transient pathological features during rapid capsule transit.

In contrast, Bluetooth Low Energy (BLE) and Wi-Fi are widely adopted wireless protocols offering larger bandwidth with complementary performance characteristics. BLE is designed for low-power operation and can support data rates of up to 1.3 Mbps \cite{9706334}, enabling higher frame rates than sub-GHz and improved temporal resolution. Although signal attenuation at 2.4 GHz is more pronounced in tissue, this limitation can be mitigated using front-end modules (FEM), and adaptive transmission power (TXP) control to preserve energy efficiency \cite{Zhou2025Closed-Loop,11175814}. Wi-Fi provides substantially higher throughput than BLE—up to several gigabits per second \cite{Mozaffariahrar2022A}—enabling high-frame-rate imaging and reducing the likelihood of missed lesions. However, this capability comes at the cost of significantly increased power consumption, with reported Wi-Fi modules exceeding 1 W \cite{10.1007/s11276-021-02780-2}, making continuous operation impractical for battery-powered capsules.

These observations reveal a fundamental trade-off in WCE communication design: low-power protocols such as BLE struggle to sustain high-frame-rate imaging, whereas high-throughput solutions such as Wi-Fi are prohibitively energy-intensive for continuous use. Importantly, the frame-rate requirement in WCE is highly dynamic, depending on capsule transit speed and the presence of suspicious lesions.

Motivated by this challenge, we propose a hybrid BLE–Wi-Fi communication architecture for WCE that dynamically adapts transmission modes according to imaging demands, as illustrated in Fig.~\ref{fig_SystemPCB}. The capsule operates in BLE mode during routine traversal or low motility to minimize energy consumption, and seamlessly switches to Wi-Fi transmission when rapid motion or suspected lesions are encountered, enabling high-frame-rate image streaming for reliable clinical interpretation. To ensure robust operation, we further introduce a frame-boundary–synchronized switching mechanism that enables reliable and lossless transitions between communication protocols while maintaining data integrity and throughput stability.


This paper makes the following contributions. First, we systematically investigate the transmission characteristics of BLE and Wi-Fi, examining Wi-Fi performance under both access point (AP) and station (STA) modes in the 2.4 GHz and 5 GHz frequency bands. Second, we design a hybrid BLE--Wi-Fi communication system and systematically study the dynamic switching process, characterizing the behavior of power consumption and FPS across switching phases. Finally, we analyze the switching latency under different image sizes to quantify the impact of frame size on handover performance. The proposed hybrid system provides practical design guidelines and valuable insights for the development of next-generation WCE systems.

\section{Methodology}


\subsection{Transmission characteristics of BLE and Wi-Fi in simulated tissue}

BLE and Wi-Fi are two widely adopted communication protocols in the Internet of Things (IoT). However, a systematic and quantitative comparison between these two protocols remains limited in the existing literature. In this study, we conduct an extensive evaluation of BLE and Wi-Fi in terms of received signal strength indicator (RSSI), throughput (FPS), and transmitter power consumption under different propagation conditions. Through this comparison, we aim to determine whether integrating both Wi-Fi and BLE into the WCE system is necessary. In the following experiments, we used the nRF7002 development kit, which integrates an nRF5340 SoC and an nRF7002 Wi-Fi companion chip.



For the BLE experiments, the nRF5340 was powered by a Power Profiler Kit II (PPK) at 3.3 V, which sampled the supply current at 100,000 samples/s to capture instantaneous current consumption \cite{nordicsemi_ppk2_userguide_v1.0.1}. The nRF7002 Wi-Fi chip was not powered and remained inactive. To ensure a stable communication link throughout the evaluation, the transmitter first operated at its maximum TXP of 3 dBm. The receiver was not implemented on a conventional computer, as computer-based receivers do not provide direct, reliable access to accurate RSSI measurements. Instead, the nRF21540 DK platform was employed to enable direct acquisition of real-time RSSI.
To further enhance the achievable TXP while reducing the energy cost, we integrated an nRF21540 FEM to the nRF7002 DK, enabling a maximum BLE TXP of up to 20 dBm. In addition, we employed an adaptive TXP control algorithm with a target throughput of 800 kbps \cite{Zhou2025Closed-Loop, Zhou2025ThroughputOptimized, 11175814}.

In the Wi-Fi experiments, two PPK modules were used to independently supply 3.3 V to the nRF5340 and the nRF7002, respectively, while simultaneously recording their current consumption. This configuration was necessary because the nRF7002 operates as a Wi-Fi companion chip and relies on the nRF5340 for system control and data processing. Consequently, the total system power consumption was calculated as the sum of the power consumption of both chips. The Transmission Control Protocol (TCP) was selected to prevent image loss and data corruption during transmission \cite{Morohashi2025Impact}. The nRF7002 supports a maximum TXP of 20 dBm. However, due to constraints imposed by the Wi-Fi protocol stack, the TXP could not be fixed and was instead dynamically adjusted according to instantaneous channel conditions \cite{nrf7002_product_spec_v1_2}. On the receiver side, a laptop equipped with an Intel 9560 network interface controller was configured with Linux and set to Wi-Fi monitor mode. Wireshark was used to capture packets and extract the RSSI relative to the transmitter. Four experimental configurations were tested: (1) the transmitter operated as STA connected to the laptop’s hotspot at 2.4 GHz, (2) the transmitter operated as a STA at 5 GHz, (3) the transmitter acted as an AP with the laptop connected to its hotspot at 2.4 GHz, and (4) the transmitter acted as an AP at 5 GHz.

To approximate electromagnetic attenuation in human tissue, a rectangular water tank measuring 40 $\times$ 30 $\times$ 25 cm was filled with water. During the experiment, the transmitter was placed at the center of the water tank and positioned at different depths to emulate varying tissue attenuation conditions. At the same time, the receiver was located in the air, 1 m from the water tank. We measured seven sets of depth: in air and at 0, 2, 4, 6, 8, and 10 cm immersion depth in water. During the experiment, the transmitter continuously sends a 40 KB image stored in its flash memory, while the receiver reconstructs the received image, calculates the throughput, and logs the real-time RSSI values.

\subsection{Reliable data transmission during protocol switching}

When the required FPS exceeds the range of the currently active protocol, a protocol switch becomes necessary. During protocol switching, if the switching sequence is not properly constrained, data loss or image corruption may occur. For example, when BLE is transmitting a video frame and a Wi-Fi transmission request is initiated midway through the frame, forcibly terminating the BLE transmission to switch to Wi-Fi will result in the loss of the partially transmitted frame. On the receiver side, this interruption can result in missing image data or visual artifacts. Such behavior is unacceptable for medical applications such as capsule endoscopy, where each switching event may cause the loss of clinically relevant information and potentially degrade diagnostic accuracy.

\begin{figure}[t]
 \centering
 \includegraphics[width=1\linewidth]{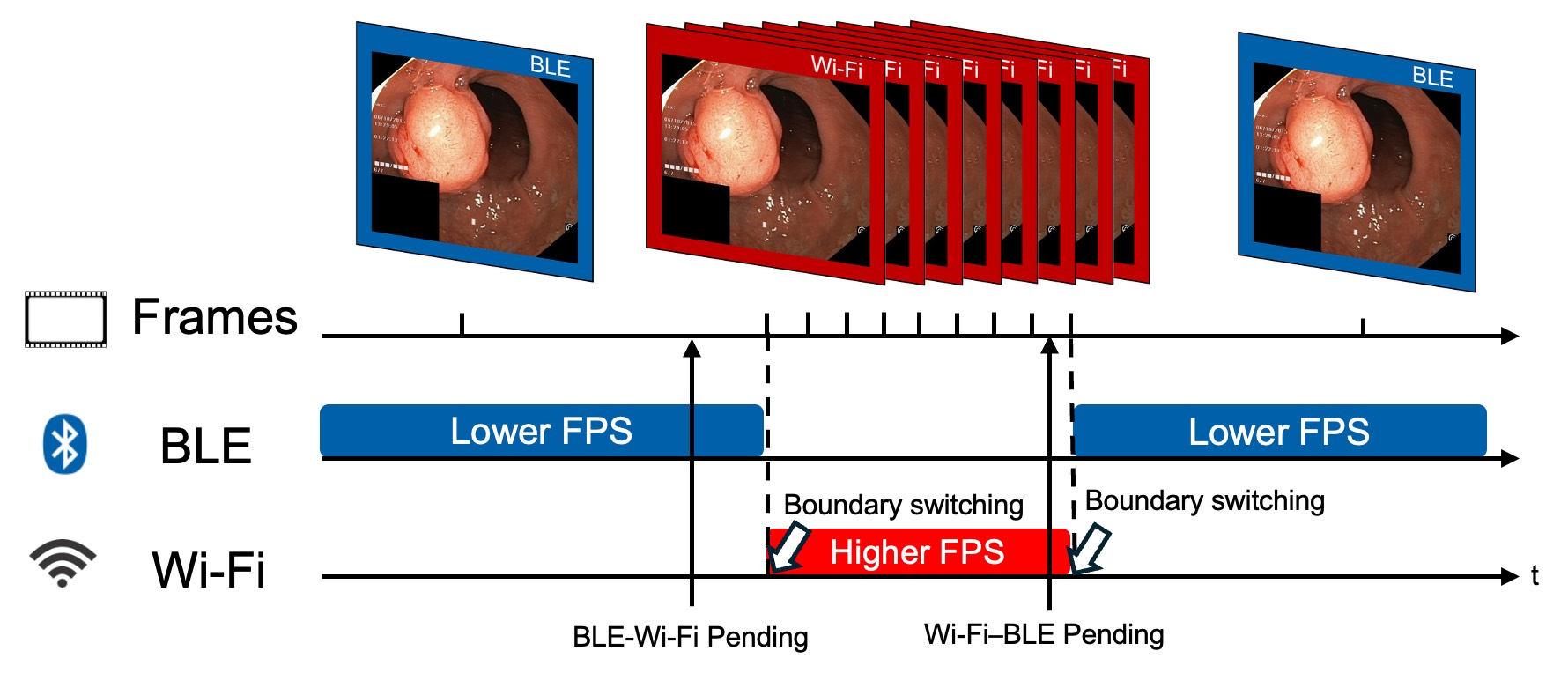}
 \caption{Bidirectional protocol switching between BLE and Wi-Fi. The system performs handovers using a frame-boundary–synchronized mechanism, ensuring lossless data transmission during protocol transitions.}
 \label{fig_Switchmech}
\end{figure}

To address this issue, we design a dedicated mechanism that supports image streaming over both Wi-Fi and BLE and enables automatic handover between the two links while preserving frame integrity. As shown in Fig.~\ref{fig_Switchmech}, the core design principle is that protocol switching is permitted only at image frame boundaries, rather than during ongoing data transmission.

When a request to initiate streaming on one protocol is issued while the other protocol is currently active, the system does not immediately perform the switch. Instead, the request is registered as a pending handover. The active link continues transmitting the current image frame at full throughput without interruption. The handover is executed only after the entire frame has been delivered, ensuring that each frame is transmitted exclusively over a single wireless protocol.

To validate the proposed switching mechanism, we conduct a complete switching experiment consisting of a BLE$\rightarrow$Wi-Fi handover followed by a Wi-Fi$\rightarrow$BLE handover. In this experiment, the system switches directly from the maximum achievable BLE frame rate (approximately 3 FPS for a 40 KB image) to the maximum Wi-Fi frame rate (approximately 25 FPS for the same image size). In this experiment, the original nRF7002 DK without an external FEM was used. Two PPK units were employed to simultaneously measure the power consumption of the nRF5340 SoC and the nRF7002 companion chip. In parallel, the receiver's FPS and the timestamps of issued switching commands are recorded synchronously.

\subsection{The relationship between image size and switching latency}

Currently, several image size specifications are commonly adopted in capsule endoscopy systems. For example, PillCam ESO2 employs an image resolution of 256 $\times$ 256, MicroCam uses 320 $\times$ 320, and OMOM adopts a higher resolution of 640 $\times$ 480 \cite{Su2025Capsule}. In this study, we aim to investigate the effectiveness of the proposed hybrid communication approach under different representative image sizes. In particular, we examine how image size influences the switching latency and efficiency during protocol handover between BLE and Wi-Fi.

Four image sizes—10, 20, 30, and 40 KB—were selected and stored in the flash memory of the nRF7002 DK for transmission. For each image size, multiple switching events were performed, including transitions from Wi-Fi to BLE and from BLE to Wi-Fi. The switching latency for transitions was recorded and analyzed to quantify the impact of image size on switching performance.

\section{Results and Discussion}
\subsection{Transmission characteristics of BLE and Wi-Fi in simulated tissue}

\begin{figure*}[t]
    \centering
    \includegraphics[width=\textwidth]{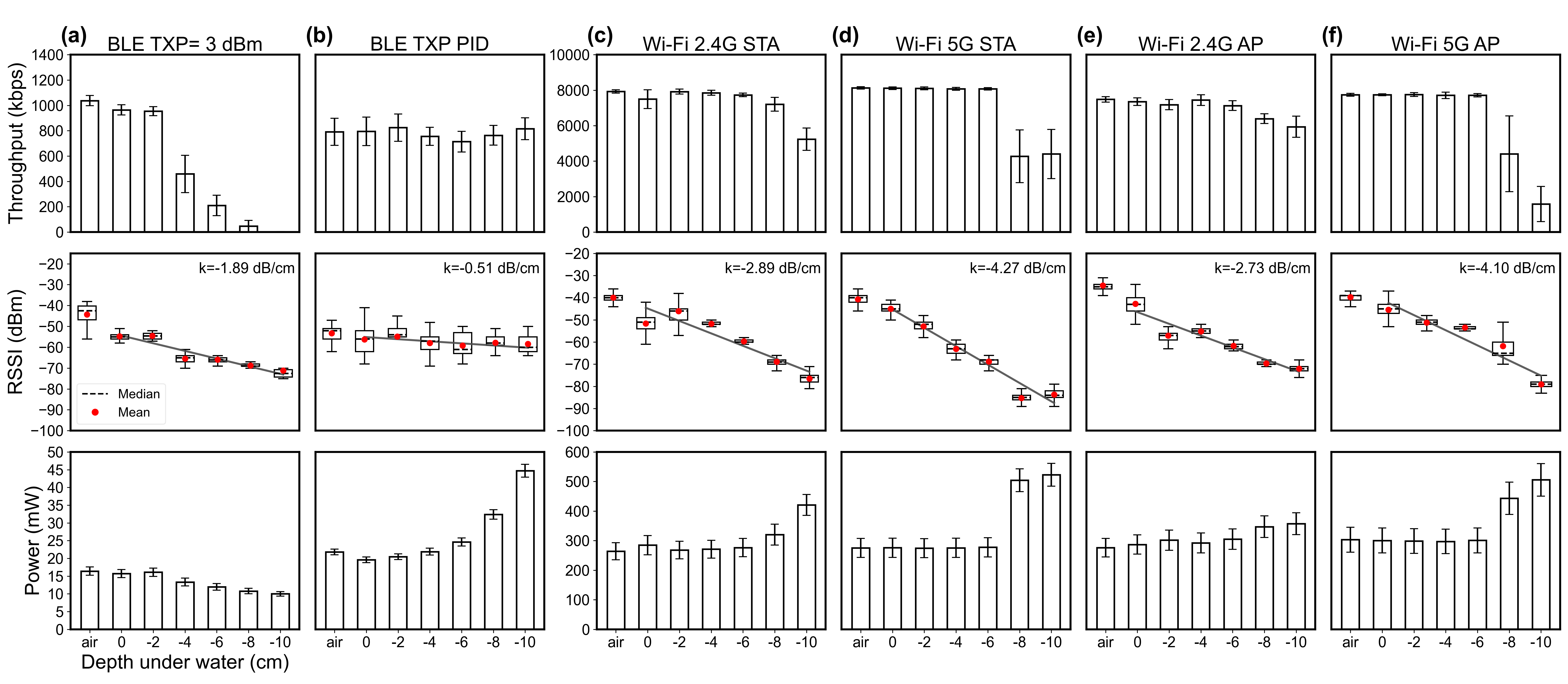}
    \caption{Comparison of BLE and Wi-Fi transmission characteristics in air and at various water depths. BLE operates at a fixed TXP of 3 dBm or adaptively, while Wi-Fi uses dynamic TXP under four configurations: 2.4 GHz STA, 2.4 GHz AP, 5 GHz STA, and 5 GHz AP.}

    \label{fig:Transmission characteristics}
\end{figure*}

As shown in Fig.~\ref{fig:Transmission characteristics}(a), for BLE communication operating at a fixed TXP of 3 dBm, the throughput decreases monotonically with increasing water depth, which can be attributed to the attenuation of electromagnetic waves in water. At shallow depths, the achievable throughput can reach approximately 1000 kbps; however, at 10 cm of water depth, it degrades significantly, dropping to only 16 kbps. A similar trend is observed for the RSSI, which decreases approximately linearly with increasing water depth, with a measured attenuation slope of -1.89 dB/cm. The measured power consumption of the BLE chip also shows a decreasing trend with increasing water depth. This counterintuitive trend arises because reduced throughput is accompanied by lower energy consumption. Under poor channel conditions, packet loss increases; packets that fail CRC checks must be retransmitted during the next connection event, introducing idle intervals and reducing radio activity. Consequently, prolonged idle time results in a lower average power consumption when throughput is severely degraded \cite{BluetoothSIG2024CoreV6}.

To mitigate such performance degradation, we integrated an FEM to extend the maximum TXP of the nRF5340 to 20 dBm. However, operating continuously at the maximum TXP is unnecessary and leads to excessive power consumption when only a predefined target throughput is required. Therefore, we adopted an adaptive TXP control mechanism, yielding the results shown in Fig.~\ref{fig:Transmission characteristics}(b). The throughput is maintained around 800 kbps with noticeable fluctuations, which arise from the inherent trade-off associated with dynamic TXP adjustment. In terms of RSSI, the signal level is regulated within a narrow range around -55 dBm, with an attenuation slope of -0.51 dB/cm. This behavior demonstrates that, under a given environment, a specific RSSI operating range corresponds to a stable target throughput. Regarding power consumption, the introduction of the FEM leads to an increase in the average power consumption as water depth increases. Compared with operation at a fixed TXP of 3 dBm at a water depth of 6 cm, the power consumption is approximately doubled. Nevertheless, the overall power consumption remains about 90\% lower than that of Wi-Fi, while still sustaining a throughput on the order of one-tenth of Wi-Fi.

For Wi-Fi transmission, a frequency-dependent comparison is conducted by examining Fig.~\ref{fig:Transmission characteristics}(c) versus (d) and (e) versus (f). The results reveal distinctly different trends between the 2.4 GHz and 5 GHz bands. At shallow water depths, the throughput at 2.4 GHz can reach approximately 8 Mbps, comparable to that at 5 GHz. However, as water depth increases, the lower frequency of the 2.4 GHz band results in lower attenuation, leading to higher throughput than the 5 GHz band at greater depths. The RSSI measurements further corroborate this attenuation behavior. Specifically, the RSSI attenuation slope for 2.4 GHz is approximately -2.8 dB/cm, whereas the slope at 5 GHz is around -4.1 dB/cm, indicating that signal attenuation at 5 GHz is roughly 1.5 times that at 2.4 GHz. In addition, power consumption measurements show that beyond a water depth of 8 cm, the increase in power consumption at 5 GHz is significantly greater than that observed at 2.4 GHz, further highlighting the disadvantages of high-frequency operation in deeper tissue-mimicking environments. The reason why Wi-Fi power has increased is that Wi-Fi employs Transmit Power Control (TPC) to compensate for signal attenuation\cite{Grilo2003Link-adaptation, Qiao2007Interference}

By comparing the AP and STA operating modes across the two frequency bands, which can be observed from Fig.~\ref{fig:Transmission characteristics}(c) and (e), as well as Fig.~\ref{fig:Transmission characteristics}(d) and (f), that both modes exhibit similar trends in throughput, RSSI, and power variation. However, as summarized in Table~\ref{tab:standby_power}, a significant difference in power consumption is observed under the standby condition, where the system remains fully initialized and ready to initiate image streaming immediately. When the transmitter is in the standby state, the STA mode consumes approximately 24 mW, whereas the AP mode requires up to 220 mW, indicating a clear advantage of the STA mode in standby power efficiency. Based on these experimental results, the STA mode operating at 2.4 GHz was selected for system design, as this configuration consistently achieved the highest and most stable throughput across all tested scenarios while maintaining the lowest overall power consumption.

At an image size of 40 KB, BLE can reliably support low-frame-rate transmission on the order of 0--3 FPS, whereas Wi-Fi can handle a significantly higher range up to 25 FPS. These complementary characteristics underscore the need for a hybrid communication system that combines the high-throughput, high-power capabilities of Wi-Fi with the low-power, low-throughput capabilities of BLE, thereby enabling efficient and reliable communication in dynamic operating environments.

\begin{table}[t]
\centering
\setlength{\tabcolsep}{4pt}
\renewcommand{\arraystretch}{0.95}
\caption{Standby power consumption under different operating modes and frequency bands}
\label{tab:standby_power}
\begin{tabular}{@{}lcccc@{}}
\toprule
 & \textbf{2.4 GHz STA} & \textbf{5 GHz STA} & \textbf{2.4 GHz AP} & \textbf{5 GHz AP} \\
\midrule
\textbf{AVG (mW)} & 25.01 & 23.03 & 218.10 & 221.36 \\
\textbf{STD (mW)} & 0.08  & 0.02  & 0.44   & 0.16   \\
\bottomrule
\end{tabular}
\end{table}

\subsection{Reliable data transmission during protocol switching}

\begin{figure}[t]
 \centering
\includegraphics[width=1\linewidth]{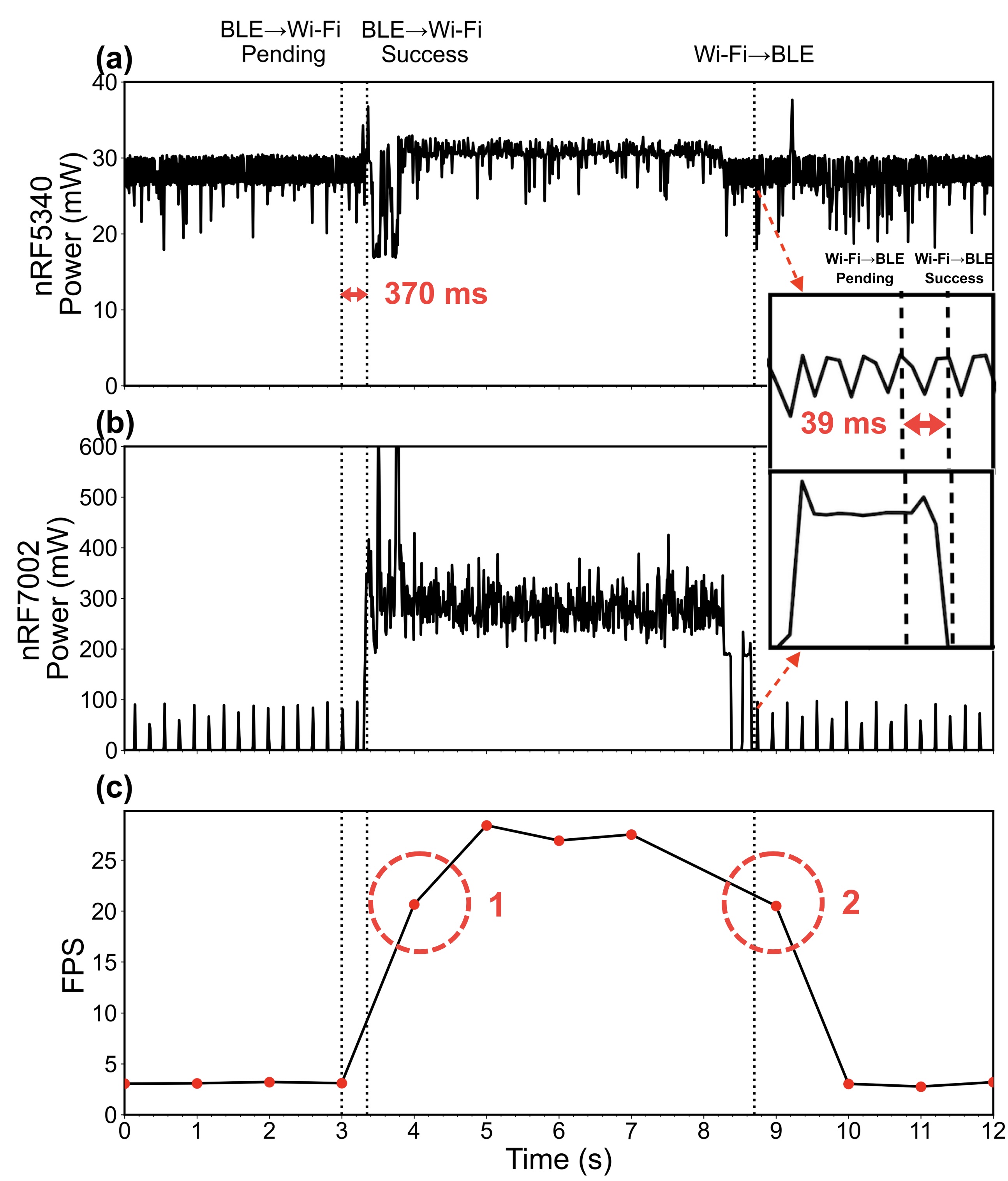}
\caption{Time-aligned measurements of power consumption and frame rate during BLE$\rightarrow$Wi-Fi and Wi-Fi$\rightarrow$BLE switching. The proposed frame-synchronized handover mechanism enables stable FPS and prevents data loss.}
\label{fig_Switch}
\end{figure}

The preceding results characterize the transmission properties of BLE and Wi-Fi when operating independently, demonstrating their respective advantages for WCE applications. However, when the required FPS exceeds the range of the currently active protocol, a protocol switch becomes necessary. Under such conditions, the system enters a critical switching state, where maintaining stable data flow and ensuring data integrity are essential. Fig.~\ref{fig_Switch} illustrates a complete switching sequence, including a transition from BLE$\rightarrow$Wi-Fi followed by a subsequent transition from Wi-Fi$\rightarrow$BLE.

Figures~\ref{fig_Switch} (a), (b), and (c) show the time-aligned evolution of the nRF5340 power consumption, the nRF7002 power consumption, and the receiver FPS during the switching process, respectively. All three traces share a common absolute time axis. We record the latency from the issuance of the BLE$\rightarrow$Wi-Fi switching command to the completion of the handover, as well as the latency from the Wi-Fi$\rightarrow$BLE command to successful switching. In this experiment, the BLE$\rightarrow$Wi-Fi switching latency is approximately 370.35ms, whereas the Wi-Fi$\rightarrow$BLE switching latency is about 39.45ms.

It can be observed that switching from BLE$\rightarrow$Wi-Fi requires a longer latency, whereas switching from Wi-Fi$\rightarrow$BLE is significantly faster. This asymmetry arises from the substantially higher throughput of Wi-Fi compared to BLE, which is consistent with the transmission characteristics shown in Fig.~\ref{fig:Transmission characteristics}. For a given amount of remaining data in a frame, the Wi-Fi$\rightarrow$BLE transition can complete much more rapidly than the BLE$\rightarrow$Wi-Fi transition, as the residual data can be drained quickly over the Wi-Fi link before handover.

Another observation is that the power transitions of the nRF5340 and the nRF7002 occur synchronously at the successful handover instant, indicating that once the proposed mechanism authorizes a BLE--Wi-Fi transition, both protocol stacks are switched concurrently. In addition, the power consumption of the nRF5340 exhibits only minor variation between BLE streaming and Wi-Fi streaming. During BLE streaming, the dominant power consumption is from BLE data transmission, whereas during Wi-Fi streaming, the primary contributor is the high-speed operation of the QSPI interface. In contrast, the nRF7002 maintains a very low standby power consumption during BLE streaming, on the order of 3.3 ± 0.37 mW, highlighting its energy efficiency when Wi-Fi transmission is inactive.

With respect to the FPS, for point 1, the FPS is calculated after the BLE$\rightarrow$Wi-Fi handover completes. Specifically, the portion of data preceding this "BLE$\rightarrow$Wi-Fi Success" is transmitted using BLE, while the portion following the "BLE$\rightarrow$Wi-Fi Success" line is transmitted using Wi-Fi. As a result, the computed FPS at point 1 lies between the steady-state FPS of BLE and that of Wi-Fi. Similarly, for point 2, the FPS is evaluated after the successful Wi-Fi$\rightarrow$BLE handover, yielding an FPS that again falls between the steady-state FPSs of the two protocols. Throughout the experiment, the receiver did not observe any incomplete frames or image artifacts, and a stable frame rate was maintained during switching. This proposed method enables synchronized protocol switching with guaranteed data integrity, ensuring stable and lossless data transmission during handover.

\subsection{The relationship between image size and switching latency}

\begin{figure}[t]
 \centering
 \includegraphics[width=1\linewidth]{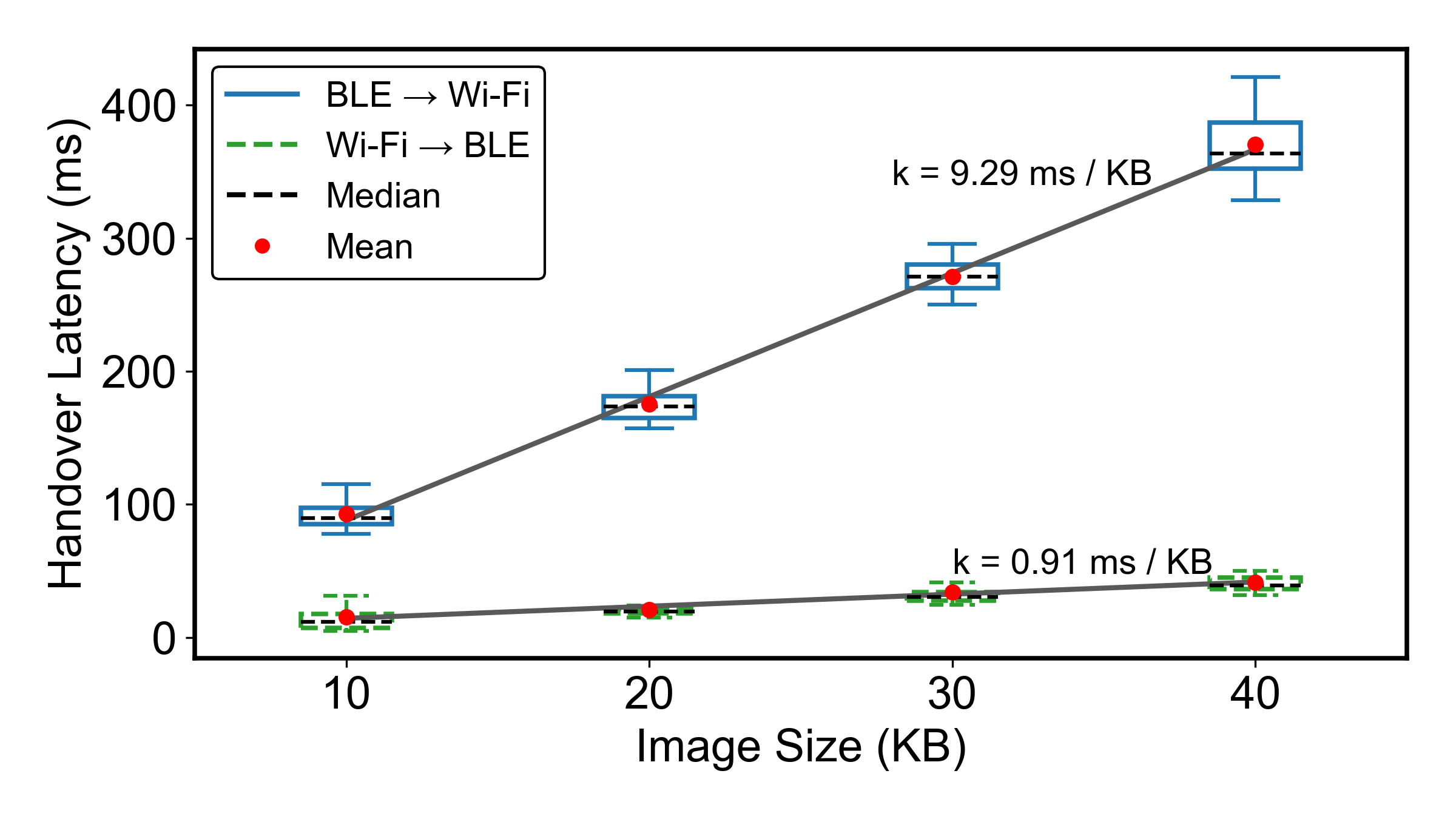}
 \caption{Switching latency between Wi-Fi and BLE under different image sizes. The measured delays include transitions from BLE$\rightarrow$Wi-Fi and from Wi-Fi$\rightarrow$BLE for image sizes of 10 - 40 KB.}
 \label{fig:switching_latency_imagesize}
\end{figure}

The relationship between image size and switching latency, obtained from multiple measurements, is illustrated in Fig.~\ref{fig:switching_latency_imagesize}. As shown in the figure, the switching latency increases monotonically with image size for both switching directions. The frame-boundary–based switching mechanism can explain this behavior.

As shown in Fig.~\ref{fig_Switchmech}, for a given image frame with a small image size, when a switching decision is made at an arbitrary time instant, the remaining portion of the frame to be transmitted is relatively small. Consequently, regardless of whether BLE or Wi-Fi is used, the residual data can be transmitted quickly, resulting in a short switching latency. In contrast, for larger image sizes, a substantial amount of data may remain to be transmitted when a switching pending is issued. Since the system must wait until the entire frame is fully delivered before completing the handover, the switching latency increases accordingly with image size.

Furthermore, by comparing the two switching directions, the growth rate of switching latency with increasing image size differs markedly between them. The measured slope for BLE$\rightarrow$Wi-Fi switching is 9.29 ms/KB, whereas that for Wi-Fi$\rightarrow$BLE switching is only 0.91 ms/KB. This order-of-magnitude difference quantitatively reflects the fact that the effective throughput of Wi-Fi is around ten times higher than that of BLE.

It is worth noting that the switching latency closely approximates the time required to transmit a complete image frame. For example, at an image size of 40 KB, the time required to transmit a single frame using BLE is approximately 320 ms, whereas the measured BLE$\rightarrow$Wi-Fi switching latency is  370.35 ± 22.15 ms. Similarly, transmitting a full 40 KB frame over Wi-Fi takes approximately 40 ms, while the Wi-Fi$\rightarrow$BLE switching latency is measured to be 41.45 ± 7.55 ms. In both cases, the observed switching latency is slightly longer than the pure frame transmission time. This additional delay is likely attributable to the overhead introduced by the frame-boundary–synchronized logic, as well as the extra computational cost associated with switching.

\section{Conclusion}

This work presents a hybrid BLE--Wi-Fi communication system for WCE that enables adaptive switching between low-frame-rate and high-frame-rate transmission modes to reduce energy consumption. A comprehensive experimental evaluation under tissue-mimicking conditions quantitatively characterizes the complementary transmission behaviors of BLE and Wi-Fi in terms of throughput, RSSI, and power consumption, thereby demonstrating the necessity of integrating both technologies.

To support practical hybrid operation, a frame-boundary–synchronized switching mechanism is proposed to ensure reliable and lossless handover between heterogeneous communication protocols. Experimental results confirm that the proposed mechanism achieves stable FPS and prevents image data loss during BLE$\leftrightarrow$Wi-Fi transitions. In addition, the impact of image size on switching latency is systematically analyzed, demonstrating that switching delay is closely related to frame transmission time and increases with image size.

Overall, the proposed hybrid BLE--Wi-Fi architecture provides an effective and energy-efficient communication solution for WCE systems operating under dynamic conditions. By leveraging the complementary advantages of BLE and Wi-Fi, this work offers practical design guidelines for next-generation capsule endoscopy platforms.

\bibliographystyle{ieeetr}

\bibliography{ref} 

\end{document}